\documentclass{article}%
\usepackage{amssymb}
\usepackage{amsfonts}
\usepackage{amsmath}%
\usepackage{graphicx}
\providecommand{\U}[1]{\protect\rule{.1in}{.1in}}

\begin{document}

\title{A competitive multi-agent model of interbank payment systems}
\author{Marco Galbiati$\dagger$ and Kimmo Soram\"{a}ki$\ddagger\bigskip$}
\date{14 May 2007}
\maketitle

\begin{abstract}
We develop a dynamic multi-agent model of an interbank payment system where
banks choose their level of available funds on the basis of private payoff
maximisation. The model consists of the repetition of a simultaneous move
stage game with incomplete information, incomplete monitoring, and stochastic
payoffs. Adaptation takes place with bayesian updating, with banks maximizing
immediate payoffs. We carry out numerical simulations to solve the model and
investigate two special scenarios: an operational incident and exogenous
throughput guidelines for payment submission. We find that the demand for
intraday credit is an S-shaped function of the cost ratio between intraday
credit costs and the costs associated with delaying payments. We also find
that the demand for liquidity is increased both under operational incidents
and in the presence of effective throughput guidelines.\bigskip\bigskip
\bigskip\bigskip\bigskip\bigskip\bigskip\bigskip\bigskip\bigskip
\bigskip\bigskip\bigskip\bigskip\bigskip

$\dagger$ Bank of England. E-mail: Marco.Galbiati@bankofengland.co.uk.\ The
views expressed in this paper are those of the authors, and not necessarily
those of the Bank of England.

$\ddagger$ Helsinki Univ. of Technology. E-mail: kimmo@soramaki.net.

The authors thank W. Beyeler, M. Manning, S. Millard, M. Willison for
important insights. Useful comments came from the participants of a number of
seminars: FS seminar (Bank of England, May 3, 2007), Central Bank Policy
Workshop (Basel, March 13, 2007), University of Liverpool, Seimnar at the
Computer Science Department (Liverpool, April 24, 2007). The usual disclaimer applies.

\end{abstract}

\section{Introduction}

Virtually all economic activity is facilitated by transfers of claims towards
public or private financial institutions. The settlement of claims between
banks takes to a large extent place at the central bank, in central bank
money. These interbank payment systems transfer vast amounts of funds, and
their smooth operation is critical for the functioning of the whole financial
system. In 2004, the annual value of interbank payments made in the European
TARGET was around \$552 trillion, in the US Fedwire system \$470 trillion, and
in the UK CHAPS \$59 trillion - tens of times the value of their respective
gross domestic products (BIS 2006). These transfers originate from customer
requests, and from the banks' proprietary operations in e.g. foreign exchange,
securities and the interbank money market. The sheer size of these transfers,
and their centrality for the functioning of a number of markets, make the
mechanisms that regulate these fluxes and the incentives that generate them
interesting to policy makers and regulators.

At present most payment systems work on a real-time gross settlement (RTGS) or
equivalent modality. In RTGS payments are settled continuously and
individually throughout the day with immediate finality. To cover the payments
banks generally use their reserve balances, access intraday credit from the
central bank or use incoming funds from payments from other banks. The first
two sources carry an (opportunity) cost which gives banks incentives to
economize on their use. We call these funds liquidity. The third source, on
the other hand, is dependent by the liquidity decisions of other banks. The
less liquidity a bank commits for settlement, the more dependent it is from
incoming payments - and may thus need to delay its own payments until these
funds arrive, causing the receivers of its payments to receive funds later. If
also delays are costly, each bank faces a trade-off between liquidity costs
and delay costs. Both aspects are dependent on the banks own liquidity
decision, but the latter is also dependent on the liquidity decisions by other banks.

This paper develops a dynamic model to study this trade-off. The model
consists of a sequence of independent settlement days where a set of
homogenous banks make payments to each other. Each of these days is a
simultaneous-move game (or a stage game) in which banks choose their level of
liquidity for payment processing. At the end of the day they receive a
stochastic payoff determined by the amount of liquidity they committed and
delays they experienced. Due to the nature of the settlement process, the
payoff function is a random variable unknown to the banks. In this context, a
reasonable assumption is that banks use heuristic, bounded-rational like rules
to adapt their behaviour over time. Hence, we simulate a learning process
taking place over many days, until banks settle down in equilibria. We are
interested in the properties of the equilibria in aggregate terms, i.e. in the
behaviour of the system as the product of independent, single agents' private
payoff maximization.

Given its game-theoretic approach, this paper is related to recent work by
Angelini (1998), Bech and Garratt (2003, 2006), Buckle and Campbell (2003) and
Willison (2004). These study various "liquidity management games" with few
(typically, two) agents and few (typically, three) periods. There, however,
the payoff function is common knowledge. Due to the complex mechanics taking
place in real payment systems this is likely to be unrealistic. Recent work by
Beyeler \textit{et al.} (2007) on the relationship between instruction arrival
and payment settlement in a similar setting shows that with low liquidity,
payment settlement gets coupled across the network and is governed by the
dynamics of the queue - and largely unpredictable when a large number of
payments are made. The present paper makes an effort to model this complexity;
in a similar spirit, it also considers a large number of banks, which settle
payments in a continuous-time day, and which interact over a long sequence of
settlement days.

Recently, a growing literature has used simulation techniques to investigate
the effects e.g. of failures in complex payment systems (see eg. BoE (2004),
Leinonen (2005), Devriese and Mitchell (2005)). These studies generally use
historical payment data and simulate banks' risk exposures under alternative
scenarios, or ways to improve liquidity efficiency of the systems. The
shortcoming of this approach has been that the behaviour of banks is not
endogenously determined. It is either assumed to remain unchanged or to change
in a predetermined manner.

The present paper tries to overcome some of the shortcomings of both "game
theoretic" and "simulation" approaches by modelling banks as learning agents.
Agents who learn about each others' actions through repeated interaction is a
recurring theme in evolutionary game theory. In one strand of the
literature\footnote{E.g. fictitious play, following Brown (1951)} the agents
know their payoff function, and learn about others' behaviour. They do so
playing the stage game repeatedly, while choosing their actions on the basis
of adaptive rules of the type "choose a best reply to the current strategy
profile" or "choose a best reply to the next expected strategy profile".
Results obtained in this strand cannot be immediately applied here: banks
cannot choose best replies as they do not know their payoff function. A second
research line does not require knowledge of the payoff function on the part of
the learners; they are instead of the kind "adopt more frequently an action
that has produced a high payoff in the past". The main results of this
literature are about the convergence (or non-convergence) of actions to
equilibria of the stage game.

The approach adopted here is close to the latter. However, because the payoffs
are calculated on the basis of a settlement algorithm, we cannot analytically
calculate the equilibria ex-ante, and then demonstrate convergence (or the
lack of it). Instead, we show convergence by means of simulations, inferring
then that the attraction points are equilibria of the stage game - in a sense
that we make precise. Because the payoff function is stochastic and unknown,
the problem of each optimizing bank lends itself to a heuristic approach. From
this perspective, our work bears strong links to the reinforcement learning
literature\footnote{See Sutton and Barto (1998) for an overview. For
Q-Learning, a common reinforcement learning technique, see Watkins and Dayan
(1992).}. From an individual agent's perspective it relates it relates to
operations research, where a typical problem is that of maximizing an unknown
function. However, in our setting the environment is not static: through time,
actions yield different payoffs both because the payoff function is random,
and because the other agents change their behaviour.

The model is rich enough to investigate a number of policy issues; here, we
focus on the aggregate liquidity of the system. As a first result we derive a
liquidity demand function, relating total funds to the ratio of delay to
liquidity costs. This function is found to be increasing to the relative cost
of delay, and S-shaped. Then, we look at the effect of operational incidents
affecting random participants of the system. We find that banks would
generally prefer to commit more liquidity in case the disruption were known -
except from the extreme cases of very low and very high delay costs.
Throughput guidelines for payment submission are a common used by
system-designers to reduce risk in payment systems;\footnote{A throughput
guideline is a constraint imposed on banks' behaviour by the system regulator;
typically, it demands that certain percentages of the total daily payments be
executed by given deadlines within the day.} we look at the effect of one such
rule on liquidity usage and find that at sufficiently low delay costs banks
would increase their liquidity holdings to contain delays. Finally, we explore
some efficiency issues, namely whether smaller systems are more or less
liquidity efficient than large ones. We find that a system with a smaller
number of banks uses less liquidity for a given level of payment activity.

The paper is organised as follows: Section 2 develops a formal description of
the model and the agents' learning process, and describes the payoff function.
Section 3 presents the results of the experiments and section 4 concludes.

\section{Description of the model}

\subsection{Stage game and its repetition\label{stage game}}

The model consists of $N\ $agents indexed by $i=1...N$, who repeatedly play a
\textbf{stage game }$\Gamma=\left\langle A,\pi_{1},\pi_{2},...\pi
_{N}\right\rangle $. Here $A=\left\{  0,1,...K\right\}  $ is the (common)
finite \textbf{action set} for each agent, and $\pi_{i}:A^{N}\rightarrow%
\mathcal{F}%
$ is $i$'s outcome function, which maps the set of action profiles into a set
of payoff distribution functions. That is, given the action profile $a\in
A^{N}$, agent $i$ receives a \textbf{stage-game payoff} drawn from a
univariate distribution $\pi_{i}\left(  a\right)  $, whose shape depends on
$N$ parameters - the stage game action profile. To keep the exposition
uncluttered, we leave the precise form of the outcome function $\pi_{i}\left(
.\right)  $ undefined at the stage. Details are given in Section
\ref{payoff function}, where we also give a precise economic interpretation to
the abstract entities introduced here. Information in the game is incomplete
as the outcome function $\pi\left(  .\right)  $ is \emph{unknown} to the
agents. Agents are risk-neutral, so they care about the expected payoff.
Hence, bank $i$ will only be concerned with its payoff functions $f_{i}\left(
a\right)  =E\left(  \pi_{i}\left(  a\right)  \right)  $.

The stage game $\Gamma$ is repeated through discrete time, running from $t=0 $
to (potentially) infinity. The action profile chosen in stage game $t$ is
denoted by $a\left(  t\right)  =\left\{  a_{1}\left(  t\right)  ,a_{2}\left(
t\right)  ,...,a_{N}\left(  t\right)  \right\}  $. A particular realization of
the payoff vector drawn from $\pi\left(  a\left(  t\right)  \right)  $, is
indicated by $y\left(  t\right)  \in%
\mathbb{R}
^{N}$, which is therefore also called the "game-$t$ payoff".

Monitoring is incomplete. At the beginning of (stage-) game $t$, each agent
$i$ knows the following: all its own previous choices and realized payoffs,
and some statistics of other's past choices $\overline{a}_{-i}\left(
k\right)  $. A (observed)\textbf{\ history} (by $i$ at time $t$) is thus
denoted by $h_{i}^{t}=\left\{  a_{i}\left(  k\right)  ,y_{i}\left(  k\right)
,\overline{a}_{-i}\left(  k\right)  \right\}  _{k=0..t-1}$. Let us call
$H^{t}$ the set of all possible histories that $i$ may observe up to $t$, and
let us define $H=\cup H^{t}$. Differently from the literature on repeated
games, but more in line with that on evolutionary game theory, we assume that
agents aim at \textbf{maximizing immediate payoffs} (instead of e.g. the
discounted stream of payoffs). That is, histories are essential to learn about
payoffs and about others' actions, but agents disregard \emph{strategic}
spillover effects between stage games. This seems a sensible assumption here:
the complexity of the environment makes it unlikely that agents anticipate all
interactions.\footnote{In the realm of reinforcement learning, immediate
payoff maximisation where actions are associated with situations is
referred\ after Barto, Sutton and Brouwer (1981) as associative search.}

\subsection{Information, learning and strategies}

Agent $i$ faces two forms of uncertainty: uncertainty about the payoff
function given others' actions, and uncertainty about other's actions. The
first element gives to our model a flavour of decision theory, the second one
is a game theory issue.

\subsubsection{Information}

As time goes by and histories are updated, agents can be seen to accumulate
information. More formally, we posit that of the whole history observed up to
$t$ each $i$ retains some multi-dimensional statistics, say $\wp_{i}\left(
t\right)  $. These are the beliefs about the state of the environment that $i$
is learning, and it constitutes the basis for the definition of strategies.
Here, $\wp_{i}\left(  t\right)  $ is composed of two parts:\smallskip

a) an "estimate" $\widetilde{f}_{i}^{t}\left(  .\right)  $ of the payoff
function $f_{i}\left(  a\right)  =$ $E\left[  y_{i}\left\vert a\right.
\right]  $;\footnote{Risk-neutral players are only interested in the expected
value of payoffs, so there is no gain in assuming that $\widetilde{f}$ instead
maps actions into payoff probability distributions.}

b) an "estimate" $\widetilde{p}_{i}^{t}\left(  .\right)  $ of probabilities
for other agents' actions in the next stage game.\smallskip

Of a whole action profile $a$, $i$ only observes its own action $a_{i}$
\emph{and} a statistic $\overline{a}_{-i}$ which correlates with $a_{-i}.$
Hence, we assume that the estimate $\widetilde{f}_{i}^{t}\left(  .\right)  $
assigns to each $\left(  a_{i},\overline{a}_{-i}\right)  $ an expected payoff.
As for the estimate b), each $i$ is assumed to maintain static expectations
about others' actions. That is, $i$ believes that $\overline{a}_{-i}$ is drawn
from a time-invariant distribution, as if other agents were adopting a
constant mixed strategy. We adopt this assumption, a classic in evolutionary
game theory, because it is simple and because it yielded the same results as
some more sophisticated forms of beliefs.\footnote{We explored in particular
the possibility that players believe that the opponents' actions follow a
Markov process. In this case, the estimate under 2) is a transition matrix,
containing the probabilities of a particular $\overline{a}_{-i}$ being played
at $t+1$, conditional on $\overline{a}=\left(  a_{i},\overline{a}_{-i}\right)
$ being observed at $t$.} Because action profiles that generate the same
statistic $\overline{a}_{-i}$ are indistinguishable to $i$, the estimate
$\widetilde{p}_{i}^{t}\left(  .\right)  $ also refers to $\overline{a}_{-i}$
instead of $a_{i}$. In the simulation, we posit that $\overline{a}_{-i}$ is
the average action of the "other" agents, which clearly takes values in
$\left[  0,K\right]  $. So, approximating to the nearest integer,
$\widetilde{p}_{i}^{t}\left(  \overline{a}_{-i}\right)  $ is a vector with $K$
entries, collecting the probabilities of any of the (other agents')
\emph{average} action being played.

\subsubsection{Learning}

The information stored in $\widetilde{f}_{i}^{t}\left(  .\right)  $ and
$\widetilde{p}_{i}^{t}\left(  .\right)  $ is updated as time goes by,
according to learning rules. A \textbf{learning rule }for agent $i$, denoted
by $\Lambda_{i}$, assigns to each observed history $h_{i}^{t-1}$ an updated
$\wp_{i}\left(  t\right)  $.

Define $I_{k}\left(  a_{i},\overline{a}_{-i}\right)  $ as the indicator
function equal to $1$ if action profile $\left(  a_{i},\overline{a}%
_{-i}\right)  $ appears at time $k$ and zero otherwise. We use the following
learning rule:%
\begin{align}
\widetilde{f}_{i}^{t}\left(  a_{i},\overline{a}_{-i}\right)   & =\frac
{\sum_{k=0...t-1}y_{i}\left(  k\right)  I_{k}\left(  a_{i},\overline{a}%
_{-i}\right)  }{\sum_{k=0...t-1}I_{k}\left(  a_{i},\overline{a}_{-i}\right)
}\\
\widetilde{p}_{i}^{t}\left(  \overline{a}_{-i}\right)   & =\frac
{1+\sum_{k=1...t-1}I_{k}\left(  \overline{a}_{-i}\right)  }{t+K}%
\end{align}

In words, $\widetilde{f}_{i}^{t}\left(  a_{i},\overline{a}_{-i}\right)  $ is
the average payoff obtained under action profile $\left(  a_{i},\overline
{a}_{-i}\right)  $ up until time $t$ excluded. Similarly, the components of
the vector $\widetilde{p}_{i}^{t}\left(  \overline{a}_{-i}\right)  $ are
calculated according to the observed frequencies, starting from an initial
estimate $1/N$. This is known as the "fictitious play" updating rule starting
from a uniform estimate; it corresponds to Bayesian updating of beliefs about
a constant, unknown distribution over the other agents' actions.\footnote{See
e.g. Fudenberg and Levine (1998) pg. 31 for details.}

\subsubsection{Strategies\label{strategies}}

A \textbf{strategy} for $i$ is a map assigning to each $\wp_{i}(t)$ an action
to be taken, i.e. some $a_{i}\left(  t\right)  $. A particular strategy can be
seen as motivated by some "rationale", resting in turn on the basis of a
learning process which we now describe.

Each $i$ is risk-neutral and aims at maximizing the expected immediate payoff.
Because $i$ believes that the opponents play a particular $\overline{a}_{-i}$
with probability $\widetilde{p}_{i}^{t}\left(  \overline{a}_{-i}\right)  $,
its strategy dictates:%
\begin{align}
a_{i}\left(  t+1\right)   & =\arg\max_{a_{i}}E\left[  \widetilde{f}_{i}%
^{t}\left(  a_{i},\overline{a}_{-i}\right)  \left\vert \widetilde{p}_{i}%
^{t}\right.  \right] \nonumber\\
& =\arg\max_{a_{i}}\sum_{\overline{a}_{-i}}\widetilde{f}_{i}^{t}\left(
a_{i},\overline{a}_{-i}\right)  \widetilde{p}_{i}^{t}\left(  \overline{a}%
_{-i}\right)  \text{\label{instant payoff maximization strategy}}%
\end{align}

The fact that banks maximize their immediate payoff is only one of the many
possible preference specifications. Alternatively, agents might also be taking
into consideration future payoffs. In this case, however, optimal strategies
would be far more complex. Indeed, discounting expected future payoffs would
create an implicit trade-off between \emph{exploitation} (the use of actions
that appear optimal in the light of the available information), and
\emph{exploration} (the use of seemingly sub-optimal actions, which might
appear such because of lack of experimentation). Our preference specification
severs this payoff-related link between stage games, which nevertheless are
interrelated because learning takes place across them. This short-sighted
maximization assumption is common in the bounded-rationality and evolutionary
game theory literature.\footnote{Fudenberg and Levine (1998) contains an
authoritative review of models with myopic agents. To quote only some of these
seminal contributions, see i) the literature on Fictitious Play by Brown
(1951), Foster et al. (1998), and Krishna et al. (1998), ii) the literature on
learning and bounded rationality inc. Kandori et al. (1993), Young (1993),
Ellison (2003), \ and Blume (1993), iii) studies on Imitation and Social
Learning by e.g. Schlag (1994).}%

\begin{figure}
[ptb]
\begin{center}
\includegraphics[
height=2.3618in,
width=3.8225in
]%
{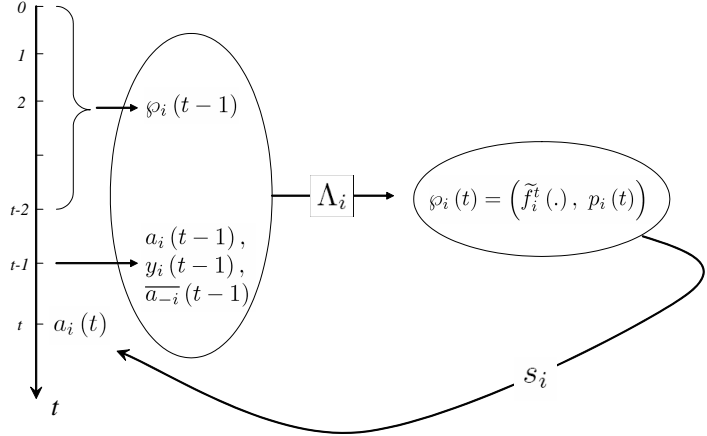}%
\caption{Information, learning and strategies}%
\end{center}
\end{figure}

Figure 1 illustrates the relationship between histories, information and
strategies. A history up to $t-2$ is summarized in $\wp_{i}(t-1)$. This, along
with the new data obtained in $t-1$, is updated into $\wp_{i}\left(  t\right)
$ by the learning rule $\Lambda_{i}$. In turn $\wp_{i}\left(  t\right)  $,
which is the information available at $t$, is mapped by a strategy $s_{i}$, to
an action $a_{i}\left(  t\right)  $.

It should be noted that in early repetitions of the stage game, $\widetilde
{p}_{i}^{t}$ is heavily influenced by the initial (arbitrary) estimate, for
which we simply use $1/K$, while $\widetilde{f}_{i}^{t}\left(  .\right)  $ is
the average of a few observed payoffs only. Hence, strict adoption of Eq.
\ref{instant payoff maximization strategy} would most likely yield, and
possibly lock into, sub-optimal actions. To avoid this, we suppose that agents
first randomly choose a certain number of actions to explore the environment,
and then start making choices as in Eq.
\ref{instant payoff maximization strategy} - which we call "informed
decisions". To ensure further exploration, each agent also tries itself out at
least once every $\overline{a}_{-i}$\ that it encounters. This models learning
from other agents.

These choices on the length of the "exploration phase" are evidently
arbitrary; however, in the model there are clearly no exploration costs, so
the length of the exploration period can indeed be assumed exogenously. On the
other hand, \emph{some} limit to exploration must be imposed, as the sheer
size of the action spaces inhibits a brute force approach, whereby $i$
collects a very large sample of all possible action profiles (and respective
payoffs) before making informed decisions.\footnote{In the simulations, each
$i$ chooses among $40$ possible actions, and $40$ are the possible average
actions by the "others" ($\overline{a}_{-i}$). Thus, full exploration would
require observing $40^{2}=1600$ different action profiles ($a_{i},\overline
{a}_{-i}$), each of which should be sampled enough times to obtain a reliable
estimate of $f\left(  .\right)  $.}

\subsection{Specification of the payoff function}

The model of learning about an unknown stochastic payoff function that is
determined party by the agent's own actions and partly by the actions of other
agents can lend itself to a number of applications. The specification of the
payoff function $f_{i}$ ties it to the problem of a payment system analyzed in
this paper.

One possible specification of $f$ \ could be a simple analytical function of
the players' actions. The problem would then become that of analyzing the
limiting behaviour of the learning rules and strategies, something that could
be done analytically, provided $f$ is simple enough. However, in quest of
increased realism in payment system modelling we specify $f$ via an algorithm
representing a "settlement day" with a large number of daily payments. To
understand why realistic analytical functions are difficult to develop,
consider the following: imagine first that banks have always enough liquidity
to make payments instantaneously. In this case payments flow undisturbed,
delay costs are zero, and only liquidity costs matter. Their calculation is
trivial. However, if banks commit less funds for settlement (as banks want to
minimize costs), it becomes more likely that the funds are at some point
insufficient for banks to execute payments immediately. As shown in Beyeler
\textit{et al.}. (2006), these liquidity shortages cause payments to occur in
"cascades", whose length and frequency bears no correlation with the
instruction arrival process that regulates payment instructions as the
settlement of payments becomes coupled across the network when incoming funds
allow the bank to release previously queued payment. As a consequence, the
flow of liquidity and thus delay times for individual banks become largely unpredictable.

In the model $a_{i}$ represents any external funding decision by the bank. The
funds allow the bank to execute payment instructions, which the bank receives
throughout the day according to a random process. Banks have costs for both
committing liquidity for settlement, and from experiencing delays in payment
processing due to insufficient funds. The settlement day is modelled in
continuous time, with time indexed as $t\in\lbrack0,T]$.\footnote{In the
previous section, $t$ indicizes "days", but we feel there is no risk of
confusion, as Section 2 and the present are relatively
independent.\label{note on t}} At any time interval $dt$, bank $i$ receives an
instruction to pay $1$ unit to any other bank $j$ with probability $\frac
{1}{N}\frac{1}{N-1}dt$. Because there are $N$ such banks $i$, and $N-1$
"other" banks $j$, the arrival of payment instructions in the whole system is
a Poisson process with parameter $1$\ so that, on the average day, $T$ payment
instructions are generated.

Payments are executed using available liquidity; $i$'s available liquidity at
time $t$ is defined as:%
\[
l_{i}\left(  t\right)  =a_{i}+\int_{s=0}^{t}\left(  y_{i}\left(  s\right)
-x_{i}\left(  s\right)  \right)  ds
\]

where $x_{i}\left(  s\right)  $ (viz. $y_{i}\left(  s\right)  $) is the amount
of $i$'s sent (viz. received) payments at time $s$. For simplicity, we assume
that every $i$ adopts the following payment rule:\footnote{The rule under
(\ref{liquidity rule}) is evidently optimal for the cost specification given
here. As banks need to pay upfront for liquidity, they have no incentive to
delay payments if liquidity is available. Under other cost specifications
(e.g. priced credit or heterogeneous payment delay costs) this would, however,
not be the case.}%
\begin{gather}
\text{at each }t\text{, execute instructions using First-in-First-out (FIFO)
as long as }l_{i}\left(  t\right)  >0\text{;}\nonumber\\
\text{else, queue received instructions}\qquad\ \ \qquad\qquad\qquad
\qquad\qquad\text{\label{liquidity rule}}%
\end{gather}

We assume that a payment instruction received by bank at $t$ and executed at
$t^{\prime}$ carries a cost equal to
\begin{equation}
CD=\kappa\frac{(t^{\prime}-t)}{T}\ \ \ \ \ \kappa>0\text{\label{delay cost}}%
\end{equation}
where $\kappa$ is the "daily interest cost" of delaying payments. Similarly,
liquidity costs (e.g. opportunity cost of collateral) are linear:
\begin{equation}
CL=\lambda a_{i},\ \ \ \ \ \lambda>0\text{\label{liquidity cost}}%
\end{equation}

Finally, the \textbf{stage game payoff} is the sum of the costs in Eq.
\ref{delay cost} and those in Eq. \ref{liquidity cost}, the former summed up
over all $i^{\prime}s$ delayed payments.\smallskip

\section{Experiments}

\subsection{Parameters and equilibria\label{payoff function}}

The continuous-time settlement day is modelled as a sequence of $10^{4}$ time
units indexed $t\in\lbrack0,10^{4}]$. Given that the arrival of payment
instructions is a Poisson process with parameter $1$, on average banks receive
a total of $10^{4}$ payment instructions per day. A sequence of days (stage
games) is called a \textbf{play}. In the simulations, we terminate a play when
no bank changes its liquidity commitment decision for $10$ consecutive days
(convergence). We run $30$ plays for each set of model parameters and find
that convergence always occurs.

Banks start the adaptation process with random decisions for liquidity, and
gradually accumulate information on the shape of the payoff function. When
enough information has been collected, banks adopt the rule described in Eq. 3
for making decisions on liquidity to commit. A series of stage games is ended
in the simulations when no bank changes its collateral posting decision for
$10$ consecutive games. This means that at this point, no bank wishes to
change its action, given the information available and given other banks'
actions.\footnote{While some changes in actions may occur due to the
randomness of payoffs and learning, these did not qualitatively change the
results in simulations with longer convergence criteria.}

Suppose that the payoff function were known by the banks; given our
specification of strategies, it would then be clear that the converged-to
actions are a Nash equilibrium of the stage game.\footnote{This simple
property of Fictitious Play stems from the fact that, if for all $t>t^{\prime
}$ the action profile is some constant $a$, then the estimates $\widetilde
{p_{i}}$ converge to the true value $\frac{1}{N}\Sigma a_{i}$. Strategies
prescribe playing a best reply to $\widetilde{p_{i}}$ so, if $a$ were not a
Nash Equilibrium, sooner or later one player would choose some $a_{i}^{\prime
}\neq a_{i}$.} We cannot quite draw the same conclusion in our setting: the
payoff exploration is necessarily partial, so it might be that some profitable
actions were never tested enough to be recognized as such. Hence, the
equilibria converged to are only "partial"-Nash equilibria, or Nash-equilibria
conditional on the "partial" information that banks have about the payoff
function. However, as we discuss later, we observe a clear consistency in
learning. This suggests that the partiality of the information collected is
sufficient, and the equilibria reached are probably good approximations of the
true stage-game Nash equilibria.

The base system consists of $15$ banks. In section 3.2 we investigate the
impact of the system size. Banks choose their action, $a_{i}$ among forty
different levels, ranging from $0$ to $80$ in intervals of $2$. The cost
functions are as in Eq. \ref{delay cost} and Eq. \ref{liquidity cost}. We
normalize liquidity costs at $\lambda=1$ and look at different values of delay
costs $\kappa$ ranging from $1/8$ to $512$ in multiples of $4$. We are
interested in the demand for liquidity (i.e. in the choices of $a_{i}$) at
different values of $\kappa$, and in the resulting settlement delays and payoffs.

\subsection{Base experiment}

As expected, with low delay costs banks tend to commit low amounts of
liquidity ($\sim$50 units) and delay payments instead, and at high cost of
delay bank prefer to commit plenty of liquidity ($\sim$1044 units) in order to
avoid expensive delays. Figures 2 and 3 illustrate typical evolution of the
simulations for two extreme levels of delay costs. The sudden changes in
liquidity correspond to the point where banks start making informed decisions
(see Section \ref{strategies}).%

\begin{figure}
[ptb]
\begin{center}
\includegraphics[
height=2.93in,
width=3.8259in
]%
{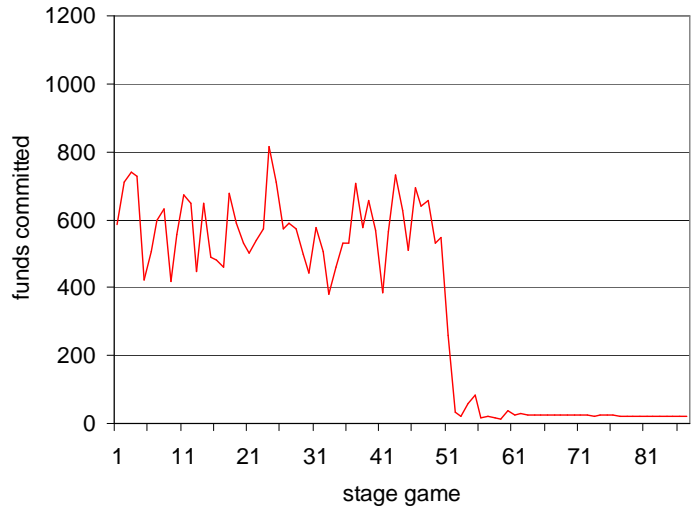}%
\caption{Total liquidity - low delay costs}%
\end{center}
\end{figure}
%

\begin{figure}
[ptb]
\begin{center}
\includegraphics[
height=2.9274in,
width=3.8225in
]%
{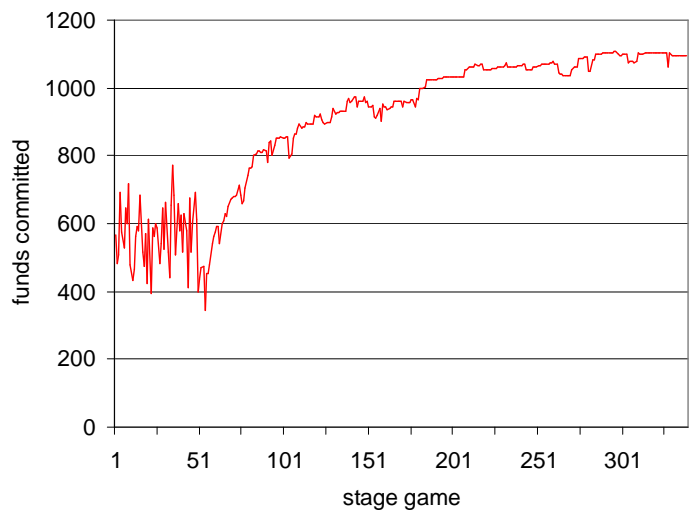}%
\caption{Total liquidity - high delay costs}%
\end{center}
\end{figure}

We find that convergence always occurs - on an aggregate level within a narrow
range. A priori, learning might be sensitive to initial observations, and
hence it might be subjected to drastic differences in the final "conclusions".
The consistency of the learning process is illustrated in Figure 4, where we
plot the converged-to value of $\sum a_{i}$ (i.e. the total liquidity
committed) across plays, for different parameter specifications. Due to
randomness - which makes histories necessarily different - the "learned"
liquidity level $\sum a_{i}$ clearly differ, but they do so within small ranges.

It should be noted that while the system consistently "learns" the same level
of \emph{total} liquidity, this can represent many configurations of single
banks' liquidity choices. Hence, our simulation don't show that always the
same equilibrium is reached; rather, that the equilibria that are reached are
characterized by a narrow span of total liquidity in the system. Given the
symmetry of the model, it is clear that for any equilibrium (i.e. any
equilibrium profile of actions $\left(  a_{1},a_{2}...a_{N}\right)  $), there
are many other equilibria obtained via a permutation of the actions between
the players that yield to same total liquidity on the system level.%

\begin{figure}
[ptb]
\begin{center}
\includegraphics[
height=2.9343in,
width=3.8225in
]%
{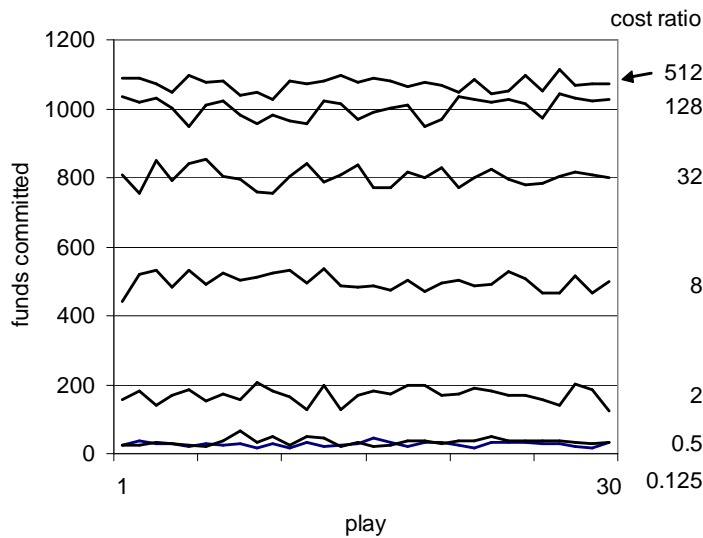}%
\caption{Total equilibrium liquidity across plays}%
\end{center}
\end{figure}

Another interesting feature of the model is its ability to match a well known
empirical fact: a low ratio of available liquidity to daily payments ("netting
ratio"), which in turn\ implies high levels of liquidity recycling. Because
the system processes on average $10.000$ payments a day, the above results
imply that the ratio in our simulation is between $0.5\%$ and $10.4\%$. For
comparison, CHAPS Sterling's netting ratio is $15\%$ (James
2004)\footnote{Calculated as the ratio of collateral used for intraday credit
to the value of payment settled.} and in Fedwire as low as $2.2\%$%
.\footnote{in 2001. Calculated as (balances + mean overdrafts) / total value.
Sources: www.federalreserve.gov/paymentsystems/fedwire} The real netting
ratios are bound to be higher due to the fact that payments in them are of
varying sizes in contrast to the more fluid unit size payments modelled here.%

\begin{figure}
[ptb]
\begin{center}
\includegraphics[
height=2.9274in,
width=3.8225in
]%
{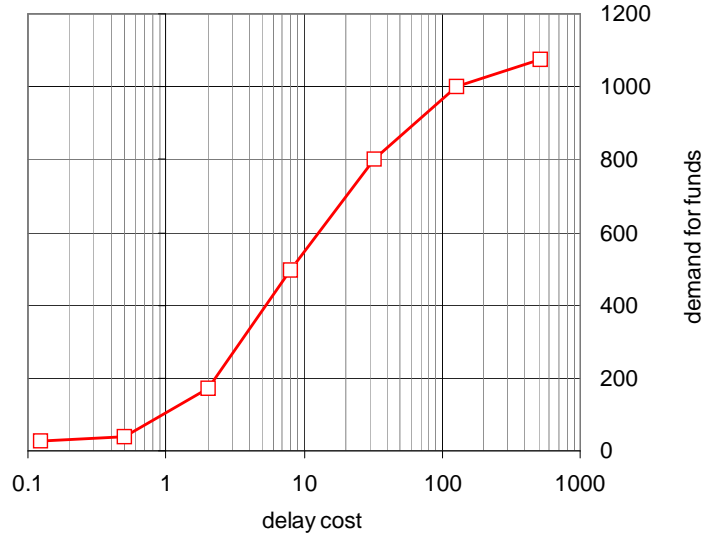}%
\caption{Liquidity as a function of delay costs}%
\end{center}
\end{figure}

Figure 5 shows the equilibrium demand for intraday credit as a function of the
cost ratio. This function is S-shaped in the exponential delay cost scale,
that is, it is relatively flat at both low and high levels of delay costs. At
comparatively \emph{low} delay costs banks evidently commit little liquidity;
hence the return for increasing liquidity holdings are \emph{high}, and so a
little more liquidity suffices to cope with increased delay costs. As a
consequence, for low delay cost levels the demand curve is flat. Consider now
the situation with high delay costs. There, the liquidity committed is high
and returns to increasing liquidity are \emph{low}, so one might think that an
increase in delay costs calls for \emph{high} extra amounts of liquidity.
However, this is not the case, because gains from liquidity indeed diminish
above a certain level when all payments can be made promptly. Hence, for high
delay costs, liquidity demand is insensitive to further increases in delay
costs, and the demand curve is flat again. In between these two extremes, the
demand for liquidity increases exponentially with delay costs.

We find that delays in the system increase exponentially as banks reduce the
amount of liquidity when this is relatively expensive compared to delaying
payments. The phenomenon is known as "deadweight losses" (Angelini 1998) or
"gridlocks" (Bech and Soram\"{a}ki 2002) in payment systems. Figure 6 shows
the relationship between system liquidity and payment delays. In intuitive
terms, the reason of this exponential pattern is the following. First, a bank
that reduces its liquidity holdings might have to delay its outgoing payments.
Second, as a consequence, the receivers of the delayed payments may in turn
need to delay their own payments, causing further downstream delays and so on.
Hence, a decrease of a unit of liquidity may cause multiple units to be
delays. Third, such a chain of delays - and hence this multiplicative effect -
is more likely and longer, the lower the liquidity possessed by the banks.
Thus the total effect of liquidity reduction acts in a compounded
(exponential) fashion.%

\begin{figure}
[ptb]
\begin{center}
\includegraphics[
height=2.9283in,
width=3.8225in
]%
{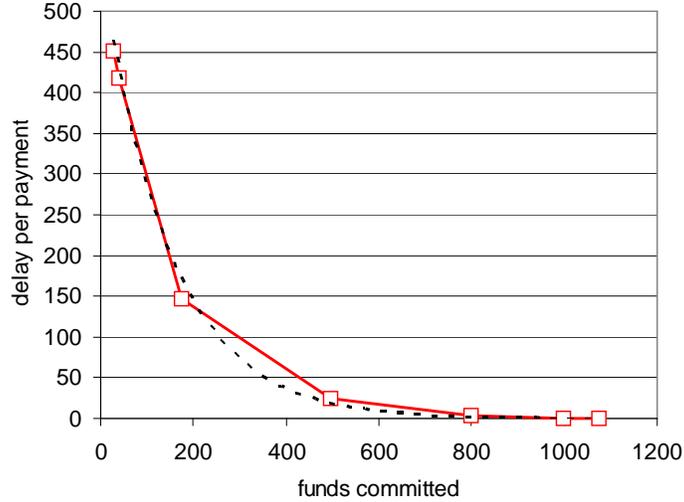}%
\caption{Delays as a function of liquidity}%
\end{center}
\end{figure}
An interesting question is how good the performance of the banks is in
absolute terms. To understand this we compare the payoffs received by the
banks through adaptation with two extreme strategies:

a) all banks delay all payments to the end of the day;

b) all banks commit enough liquidity to be able to process \emph{all} payments
promptly.\footnote{In fact the liquidity committed in the simulation with the
highest delay cost was used as the scenario for prompt payment processing.}

The comparison between the performance of these two pure strategies and the
learned strategy is shown in Figure 7. For any cost ratio, the adaptive banks
obtain better payoffs than any of the two extreme strategies - except for the
case with high liquidity costs when the costs are equal. Banks manage to learn
a convenient trade-off between delay and liquidity costs. On the contrary, the
strategy under a) becomes quickly very expensive as delay costs increase, and
the strategy under b) is exceedingly expensive when delays are not costly.%

\begin{figure}
[ptb]
\begin{center}
\includegraphics[
height=2.9352in,
width=3.8225in
]%
{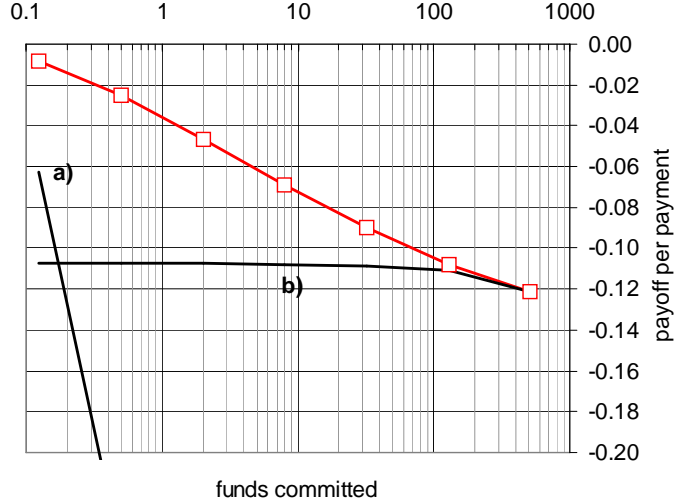}%
\caption{Comparison of strategies: a) min liquidity; b) min delays}%
\end{center}
\end{figure}

\subsection{Impact of network size}

In order to investigate the impact of system size on the results presented in
the previous section, we ran simulations varying the number of participants.
To ensure comparability, we kept the number of payments constant across
simulations and maintained the network complete.

We observe that liquidity demand increases with the system size, and the
increase is more pronounced the higher the delay costs. For example, the
demand function is unchanged at low delay cost while, for high delay costs, a
50-bank system requires $215\%$ the liquidity needed in a 15-bank system.%

\begin{figure}
[ptb]
\begin{center}
\includegraphics[
height=2.9291in,
width=3.8233in
]%
{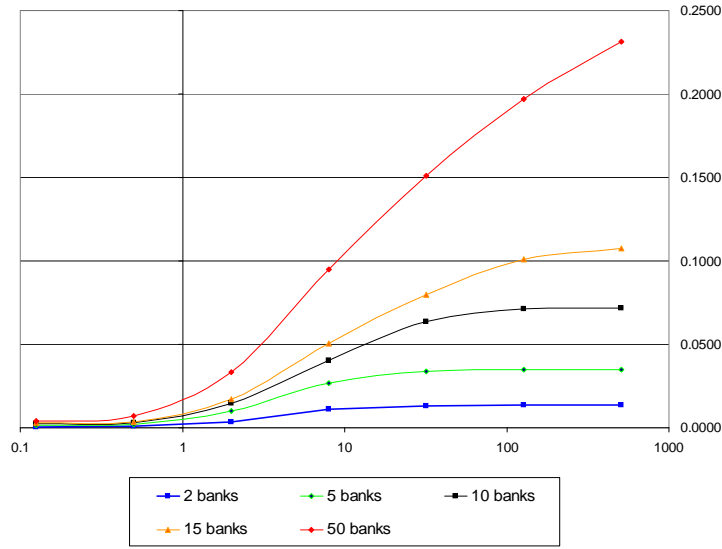}%
\caption{The effect of system size on the demand for liquidity}%
\end{center}
\end{figure}

Similar results hold about delays. The relationship between liquidity
committed and delays remains close to exponential irrespective of system size;
however, larger systems experience more delays, for any level of initial
liquidity (see Figure 9).

An intuitive explanation of these phenomena could be the following. First,
note that if the number of participants is increased by a factor $x$ (keeping
turnover constant), the volatility of the balance of each bank is multiplied
by a factor $1/x^{\prime}>1/x$ - we show this in a moment. Second, suppose
that i) the optimal $a_{i}$ is proportional to the volatility of a bank's
balance $\delta$ (i.e. $a_{i}=z\delta$)\footnote{This is exactly the case if a
bank chooses $a_{i}$ as to cover $z$ \textquotedblleft standard
deviations\textquotedblright\ from the average balance.} and that ii) banks
post all the same amount of liquidity (i.e. $a_{i}=a_{j}$). It then simply
follows that the total amount of liquidity increases with the system's size:
$\left(  nx\right)  z\frac{\delta}{x^{\prime}}>nz\delta$ (here $nx$ is the
number of banks in the larger system, and $\frac{\delta}{x^{\prime}}$ the
corresponding volatility of balances).

The key point is that, if the number of participants is increased (keeping
turnover constant), the volatility of banks balances rises more than
proportionally. To see why this is the case, consider the simplified but
illustrative situation where liquidity is abundant, so there are no delays. In
this case, a bank's net position is the sum of a series of random
perturbations (incoming and outgoing payments), equally likely to affect it
positively and negatively. In other words, a bank's net position is a random
walk, whose value after $n$ perturbations averages zero, with a standard
deviation $\sqrt{n}$. By increasing the system size by a factor $x$, the
orders are distributed over more banks, so the average number of perturbations
for any given time interval is multiplied by $1/x<1$. Accordingly, the
standard deviation of the balances at the end of any time interval is
multiplied by $1/x^{\prime}=\sqrt{1/x}>1/x$.

\subsection{Throughput guidelines}

Some interbank payment systems have guidelines on payment submission jointly
agreed upon by the system participants\footnote{E.g. the FBE (Banking
Federation of the European Union 1998) has set guidelines on the timing of
certain TARGET payments. In CHAPS Sterling, members must ensure that on
average (over a calendar month) 50\% of its daily value of payments are made
by 12 pm and that 75\%, by value, are made by 2.30 pm. (James 2004).}. The
rationale for throughput guidelines is to induce early settlement in order to
e.g. reduce operational risk or perceived coordination failures among
participant. For example, if a large chunk of payments are settled late in the
day, an operational incident would be more severe as more payments could
potentially remain unsettled before close of the payment system and the
financial markets where banks balance their end-of-day liquidity positions.

We simulate a particular realization of throughput guidelines by introducing
an additional penalty charge for delays that last longer than one tenth of the
settlement day (i.e. 1000 time units). The penalty charge is set to 64, in
order to sufficiently penalize non-compliance with the rules.

Figure 10 shows the impact of the throughput guidelines on the amount of
liquidity committed by the banks. When delay costs are high, banks already
commit enough liquidity to avoid long delays, so the throughput guidelines
have no effect. They do however, in the case of low delays costs, induce banks
to commit more liquidity. Not surprisingly, this comes at a cost to the banks,
which are forced away from their first-best choice. The increase in costs are
of the order of 70\% at the lowest level of delay costs, and 20\% at the
second lowest. The payoff comparison is shown in Figure 11.%

\begin{figure}
[ptb]
\begin{center}
\includegraphics[
height=2.9308in,
width=3.8225in
]%
{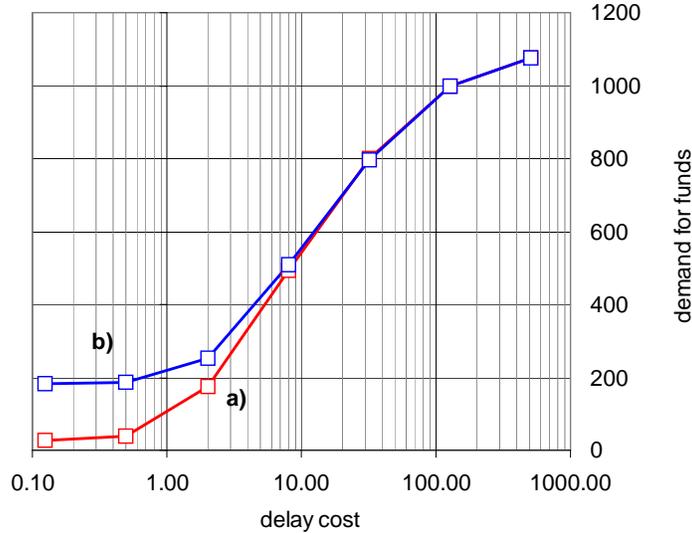}%
\caption{Impact of throughput guidelines (b) on liquidity}%
\end{center}
\end{figure}
%

\begin{figure}
[ptb]
\begin{center}
\includegraphics[
height=2.9317in,
width=3.8216in
]%
{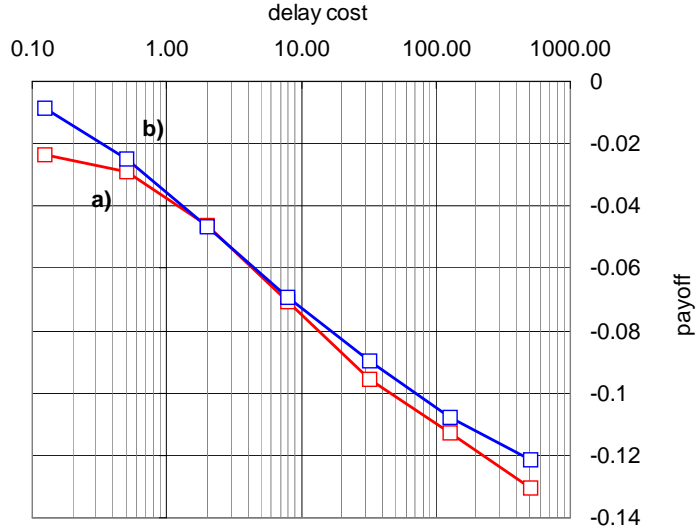}%
\caption{Impact of throughput guidelines (b) on payoffs}%
\end{center}
\end{figure}

\subsection{Operational incident}

Short term outages by banks in the payment system are rare in actual payment
systems, but do take place occasionally. In a typical scenario a participating
bank experiences problems connecting to the system due to temporal
unavailability of IT systems or telecommunication facilities. Due to the
design of the payment systems, a disconnected bank can in such situations
generally still receive payments to its account at the central bank, but
cannot submit instructions to pay from its account. Unless other banks stop
paying to the troubled bank, it quickly becomes a liquidity sink, and the
liquidity available for settling payments at other banks is
reduced.\footnote{see e.g. analysis on the impact of the 9/11 terrorist
attacks in McAndrews and Potter (2002), Lacker (2004), and Soram\"{a}ki et al
(2007)}

In this set of simulations we ask the question of how much liquidity banks
would wish to commit in such a situation, i.e. what is the impact of an
operational outage on the demand for intraday credit. The banks are assumed to
be unaware of the possible incident, and unable to discriminate among their
counterparts, so the intraday liquidity management rule under Eq.
\ref{liquidity rule} is still adopted. Under these assumptions, we simulated a
scenario where a randomly selected bank can receive, but cannot send payments
for the first half of the settlement day. On average, this means that up to
$T/2N=10.000/(2\cdot15)\simeq333$ liquidity units cannot be used by other
banks as a source of liquidity\footnote{The shortage of liquidity equals the
number of payment orders received by the distressed bank, and not yet executed
until the second half of the day.}. Depending on the delay cost, this figure
varies between 1900\% and 30\% of the average total liquidity injected in the
system at the beginning of the day (the first figure being for the case when
both delay cost and liquidity demand are low, and the second for when costs
liquidity demand are both high).

We found that the effect of operational incidents on the demand for liquidity
is highest at a relative delay/liquidity cost ratio of $2$ - hence fairly low
in the range; at this point, the increase in intraday credit demand is $144$
units, or an increase of $85\%$. It should be noted that banks do not
compensate for the full amount of liquidity "trapped" by the distressed bank,
but prefer to partly make up for that, and partly increase delays. For higher
delay costs, delays remain approximately unchanged compared to the scenario
without the incident. Finally, when delay costs are lower than liquidity costs
(i.e. for a cost ratio
$<$%
1), banks prefer to hold about the same amount of liquidity as without the
incident, and experience the delays caused by the reduced liquidity. In this
case, the impact of an operational incident increases both the demand for
liquidity (but less than what was trapped) and delays - more so the one which
is less costly.%

\begin{figure}
[ptb]
\begin{center}
\includegraphics[
height=2.9291in,
width=3.8225in
]%
{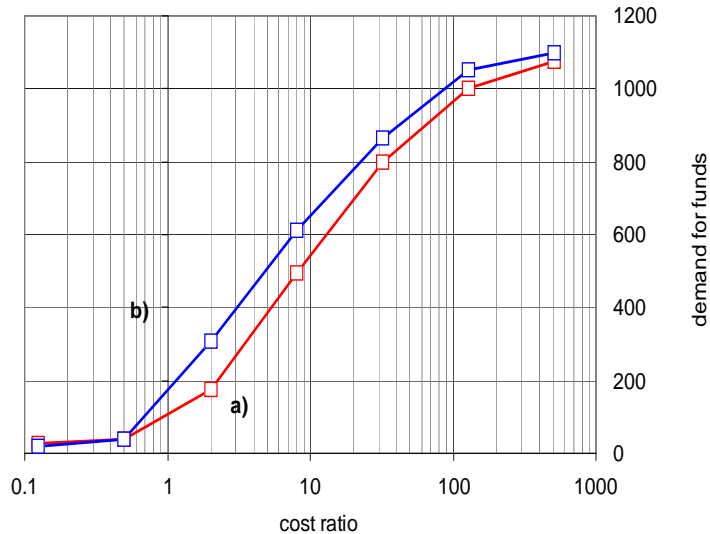}%
\caption{Liquidity - normal circumstances (a) and operational incidents (b)}%
\end{center}
\end{figure}
%

\begin{figure}
[ptb]
\begin{center}
\includegraphics[
height=2.8202in,
width=3.8225in
]%
{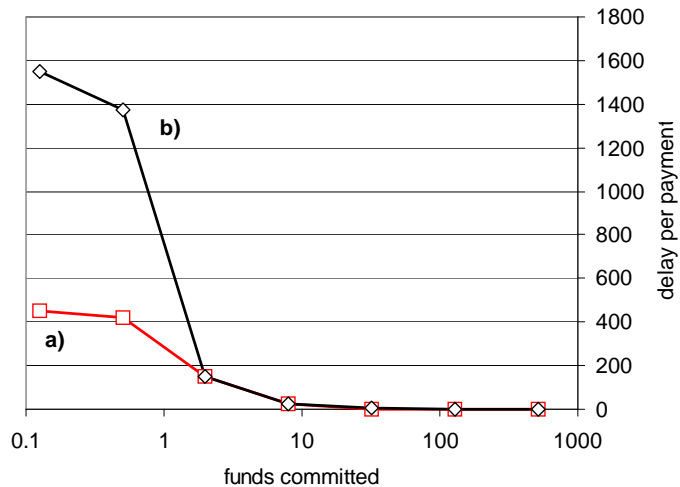}%
\caption{Delays as a function of cost ratio (b: incident)}%
\end{center}
\end{figure}

\section{Summary and conclusions}

In this paper we developed an agent-based, adaptive model of banks in a
payment system. Our main focus is on the demand for intraday credit under
alternative scenarios: i) a "benchmark" scenario, where payments flows are
determined by the initial liquidity, and by an exogenous arrival of payment
instructions; ii) a system where, in addition, throughput guidelines are
exogenously imposed iii) a system subject to operational incidents.

It is well known that the demand for intraday credit is generated by a
trade-off between the costs associated with delaying payments, and liquidity
costs. Simulating the model for different parameter values, we were able to
draw with some precision a liquidity demand function, which turns out to be is
an S-shaped function of the delay / liquidity cost ratio. We also looked at
the costs experienced by the banks, as a function of the model's parameters.
By the process of individual payoff maximization, banks adjust their demand
for liquidity up (reducing delays) when delay costs increase, and down
(increasing delays), when they rise. Interestingly however, the absolute delay
cost remains approximately constant when the ratio delay/liquidity costs
changes. As expected, the demand for intraday credit is increased by an
operational incident. However, this effect is found to be important only if
liquidity is costly compared to delaying payments. Likewise, throughput
guidelines increase the demand for intraday credit - as banks try to avoid
penalties for not adhering to them. In total this reduces the payoffs of the
banks. Nevertheless, throughput guidelines may be beneficial when additional
benefits that are not in the current model\ are taken into account (among
these, benefits related to reducing operational risk).

This model produces realistic behaviour, suggesting that it may be used to
investigate a wide array of issues in future applications. A number of
extensions are possible. First, alternative specifications for the instruction
arrival process may be applied (see e.g. Beyeler at al. (2006)).
Alternatively, one could change the assumptions on the banks' network: while
the complete network assumption implicitly adopted here fits well with e.g.
the UK CHAPS system, an interesting question is how other topologies such as a
scale free network topology such as in Fedwire (Soram\"{a}ki \textit{et al.}.
2007) would affect the results. Also, different individual preferences could
be investigated. We assumed that banks are risk neutral \emph{and} interested
in maximising their immediate payoffs; it would be interesting to verify if
the introduction of risk aversion and / or preferences over expected stream of
payoffs may change the results. Finally, more complex behaviour can be easily
studied within our model; for example, the "pay-as-much-as-you-can" rule for
queuing payments could be replaced by sender limits. Similarly, more
sophisticated strategies can be easily modelled, supposing e.g. that banks
keep constant their actions for a number of periods (to gather more data and
explore the environment), instead of exploiting after a fixed amount of time
what seems to be the best action.

\bibliographystyle{boewp}
\bibliography{SuppliedRefs}

\end{document}